\documentclass[twocolumn,prb,notitlepage]{revtex4-1} 

\usepackage{amsmath}  
\usepackage{amsfonts} 
\usepackage{graphicx} 
\usepackage{hyperref}

\newcommand{\eq}[1]{Eq.\,\eqref{#1}}
\newcommand{\eqs}[1]{Eqs.\,\eqref{#1}}
\newcommand{\cc}{\mathrm{c}}
\newcommand{\co}{\mathrm{const.}}
\newcommand{\ud}{\mathrm{d}}
\newcommand{\G}{\mathrm{G}}
\newcommand{\s}[1]{{\text{\tiny $#1$ }}\hspace{-2pt}}

\newcommand{\dpart}[2]{\frac{\partial #1}{\partial  #2}}

\newcommand{\8}{\s{\infty}}

\begin{document}

\title{Incompressible wind accretion}

\author{Emilio Tejeda}
\email{etejeda@astro.unam.mx} 
\affiliation{Instituto de Astronom\'{i}a, Universidad Nacional Aut\'{o}noma de M\'{e}xico, AP 70-263, Ciudad de M\'exico, 04510, Mexico}

\date{\today}

\begin{abstract}
We present a simple, analytic model for the accretion flow of an incompressible wind onto a gravitating object. This solution corresponds to the Newtonian limit of a previously known relativistic model for a fluid obeying a stiff equation of state for which the sound speed is constant everywhere and equal to the speed of light. The new solution should be useful as a benchmark test for numerical hydrodynamics codes and, moreover, it can be used as an illustrative example in a gas dynamics course.
\end{abstract}

\maketitle

\section{Introduction}
\label{Sec1}

The study of wind accretion phenomena has been an active field of research since the pioneering work of Hoyle \& Lyttleton\cite{hoylely} and Bondi \& Hoyle.\cite{hoyle} In its basic formulation, this problem deals with the accretion flow onto a massive, gravitating object traveling at a constant velocity across an otherwise unperturbed, homogeneous gaseous medium. Alternatively, this same situation can be reversed to consider instead a constant wind accreting onto a massive object held fixed at the origin of coordinates. Hoyle \& Lyttleton provided a semi-analytical description of this problem by approximating the incoming accretion flow by ballistic trajectories. Analytic solutions have been found in the spherical symmetric case in which the relative velocity between the central object and the surrounding medium is zero.\cite{bondi52,michel72}  Advances for the more general case of a non-zero velocity have been mostly based on numerical studies.\cite{hunt71,bisnovatyi,petrich89, matsuda91,font98a,foglizzo05,penner11,zanotti11,cruz12,blondin12,LG13} For general reviews on wind accretion see, e.g. Livio\cite{livio91} and Edgar.\cite{edgar04}

The only full analytic solution to a wind accretion problem has been that found by Petrich, Shapiro \& Teukolsky\cite{petrich88} (PST hereafter). The PST model describes the relativistic wind accretion flow onto a black hole under the approximation of a fluid obeying a stiff equation of state for which the fluid sound speed equals the speed of light everywhere. Because of this, the PST model was thought to have no  Newtonian analogue. In this article, however, we present a simple, analytic solution for an incompressible wind accreting onto a Newtonian gravitating object and show that it corresponds to the non-relativistic limit of the PST solution. Besides filling this blank in the literature, this simple solution constitutes in itself an interesting example that can be discussed as part of a gas dynamics course. The incompressible flow approximation is valid whenever pressure gradients lead to a negligible variation in fluid density. Although this approximation has a rather limited applicability in astrophysics (with the possible exception of the dense interior of a neutron star), the analytic model presented in this article should be useful as a test solution for benchmarking numerical codes.
 
\section{Analytic Solution}
\label{Sec2}

Consider a steady wind passing by a gravitating object of mass $M$ sitting at the origin of coordinates. We assume that the gas consists of an incompressible fluid described by a constant density $\rho$ and two variables: its pressure $P$ and a velocity field $\vec{v}$. Furthermore, we take the central object as a spherically symmetric, passive sink of gas. Far away from the central object, the fluid is described by the boundary conditions
\begin{gather} 
P|_\8 = P_\8, \label{e1}\\ 
\vec{v}|_\8 = v_\8\,\hat{z}, \label{e2}
\end{gather}
where we have aligned the $z$ axis with the incoming wind direction. 

In general, the flow dynamics will be governed by the continuity and Euler equations:
\begin{gather}
\dpart{\rho}{t} + \nabla\cdot\left(\rho\,\vec{v}\right) = 0, 
\label{e3}\\
\dpart{\vec{v}}{t} + \vec{v}\cdot\nabla\vec{v} = -\frac{1}{\rho}\nabla P - \frac{\G M}{r^2}. 
\label{e4}
\end{gather}
Under the assumptions of stationarity and incompressibility, \eq{e3} reduces to
\begin{equation}
\nabla\cdot\vec{v} = 0, 
\label{e5}
\end{equation}
while integration of \eq{e4} leads to the Bernoulli constant 
\begin{equation}
B = \frac{v^2}{2} + \frac{P}{\rho} - \frac{\G M}{r} = \frac{v^2_\8}{2} + \frac{P_\8}{\rho}, 
\label{e6}
\end{equation}
where we have imposed the boundary conditions in \eqs{e1} and \eqref{e2}.

The boundary condition for the velocity field at infinity in \eq{e2} corresponds to an irrotational flow, i.e.~\mbox{$\nabla\times\vec{v}= 0$}. Given the axisymmetry of the problem, we shall expect this condition to hold everywhere else. We can then propose a velocity potential $\Phi$ such that $\vec{v}=\nabla\Phi$ which, according to \eq{e5}, satisfies the Laplace equation
\begin{equation}
\nabla^2\Phi = 0. \label{e7}
\end{equation}
Adopting spherical coordinates, an axisymmetric solution to this equation is given by
\begin{equation}
\Phi = \sum_{n=0}^\8 \left(A_n\,r^n + B_n\,r^{-(n+1)} \right) P_n(\cos\theta),\label{e8}
\end{equation}
where $A_n$ and $B_n$ are constant coefficients and $P_n(\cos\theta)$ is the Legendre polynomial of degree $n$. With these same coordinates, the boundary condition in \eq{e2} is
\begin{equation}
\vec{v}|_\8 = v_\8\,\hat{z} = v_\8\left(\cos\theta\, \hat{r} - \sin\theta\,\hat{\theta}\right),
\label{e9}
\end{equation}
from where it follows that
\begin{gather}
\left.\left(\dpart{\Phi}{r}\right)\right|_\8 = \left.\left(\frac{\ud r}{\ud t}\right)\right|_\8 = v_\8\,\cos\theta, \label{e10}\\
\left.\left(\frac{1}{r}\dpart{\Phi}{\theta}\right)\right|_\8 = \left.\left(r\,\frac{\ud \theta}{\ud t}\right)\right|_\8 = - v_\8\,\sin\theta.
\label{e11}
\end{gather}
Substituting $\Phi$ from \eq{e8} into \eqs{e10} and \eqref{e11} we obtain  
\begin{gather}
A_1 = v_\8, \label{e12}\\
A_n = 0 \qquad \text{for} \qquad n\ge2. \label{e13}
\end{gather}

On the other hand, since we are interested in finding a steady-state solution, we have the additional condition of a constant accretion rate $\dot{M}$ across any closed surface surrounding the central object. In particular, if we take a sphere of radius $r$ centered at the origin, we have 
\begin{equation}
\begin{split}
\dot{M}  & = \int_0^{2\pi} \int_0^\pi \rho\left(\vec{v}\cdot \hat{r}\right)r^2\sin\theta\,\ud\theta\,\ud\phi \\
& = 2\pi\,r^2\int_0^\pi \rho\left(-\frac{\ud r}{\ud t}\right)\sin\theta\,\ud\theta \\
& = 4\pi\rho\,B_0,
\end{split}
\label{e14}
\end{equation}
which implies that 
\begin{equation}
B_0 = \frac{\dot{M}}{4\pi\rho}.
\label{e15}
\end{equation}

Note that at this point we do not have any restriction on the coefficients $B_n$ for $n>0$ since, due to the orthogonality of the Legendre polynomials, it will always be guaranteed that
\begin{equation}
\int_0^\pi P_n(\cos\theta)\sin\theta\,\ud\theta = 0. \label{e16}
\end{equation}
These higher order multipoles can in principle be used to match some given inner boundary condition close to the accretor. In the absence of any such boundary condition, we take the lowest order solution given by
\begin{equation}
\Phi = v_\8\,r\,\cos\theta + \frac{\dot{M}}{4\pi\rho\,r}, \label{e17}
\end{equation}
which leads to the velocity field
\begin{gather}
\frac{\ud r}{\ud t} =  v_\8\left(\cos\theta - \frac{s^2}{r^2}\right)\label{e18},\\
\frac{\ud \theta}{\ud t} = - \frac{v_\8}{r}\,\sin\theta,\label{e19}
\end{gather}
where we have introduced the stream length scale
\begin{equation}
s = \sqrt{\frac{\dot{M}}{4\pi\rho\,v_\8}}.\label{e20}
\end{equation}

An equation for the streamlines can be obtained by combining \eqs{e18} and \eqref{e19}, which results in the differential equation
\begin{equation}
\frac{\ud r}{\ud \theta} = \frac{s^2\csc\theta}{r} -r\,\cot\theta. \label{e21}
\end{equation}
\eq{e21} can be integrated at once to give
\begin{equation}
r = \frac{\sqrt{b^2 - 2\,s^2(\cos\theta+1)}}{\sin\theta}, \label{e22}
\end{equation}
where $b$ is an integration constant which we have chosen in such a way that it corresponds to the impact parameter of each streamline, i.e.
\begin{equation}
\left.\left(r\,\sin\theta\right)\right|_{\theta=\pi} = b.  \label{e23}
\end{equation}

\begin{figure*}
\begin{center}
  \includegraphics[width=0.45\textwidth]{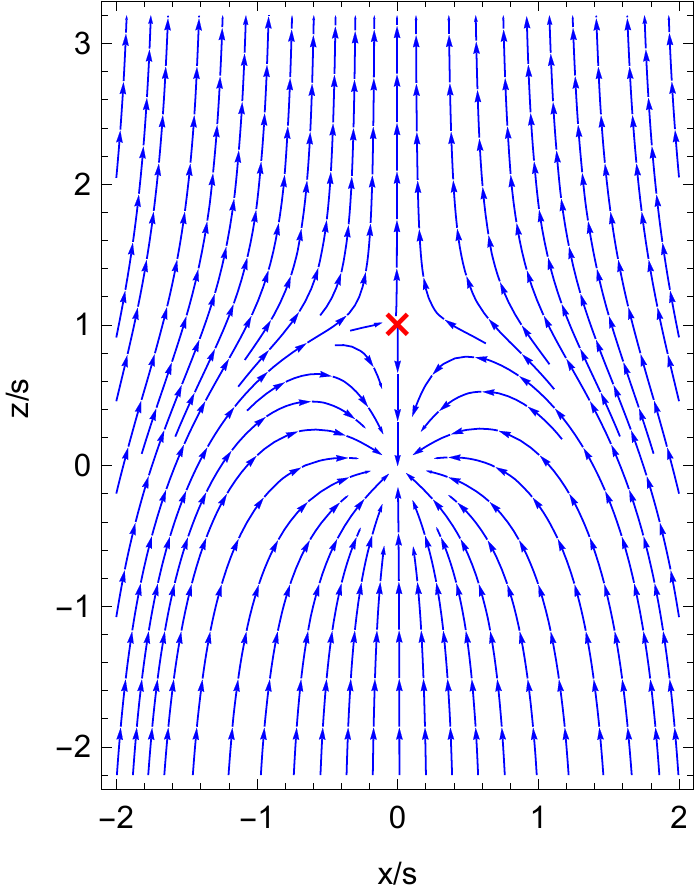}   
  \includegraphics[width=0.45\textwidth]{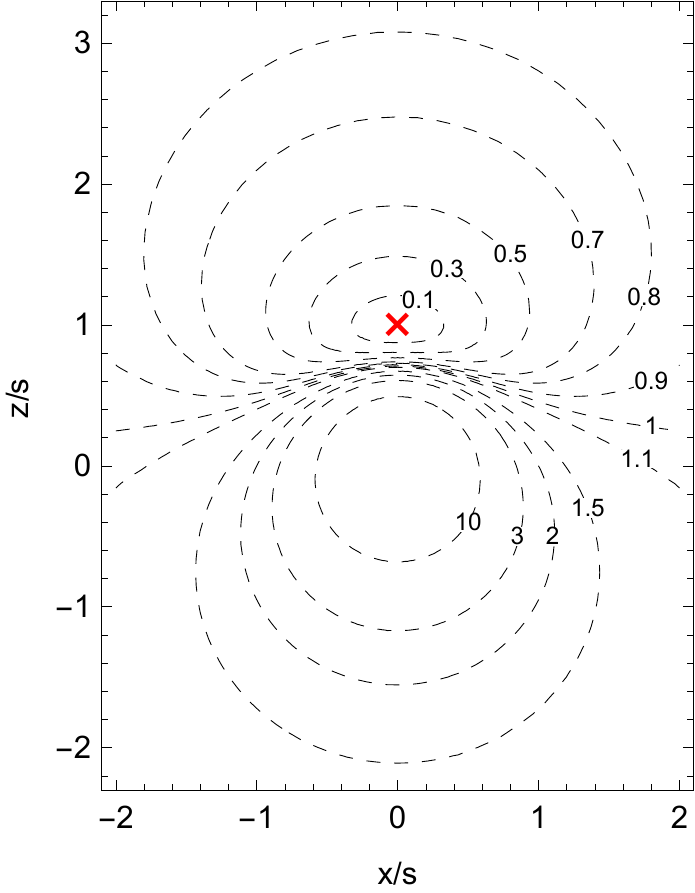}
\end{center}
 \caption{Wind accretion of an incompressible fluid. The left-hand panel shows the streamlines as described by \eq{e22}. The right-hand panel shows the isocontour levels of the magnitude of the velocity field in \eqs{e18} and \eqref{e19}, the contour labels being expressed in units of $v_\8$. The stagnation point is marked with a red cross in both panels.}
\label{f1}
\end{figure*}

In Figure~\ref{f1} we show the streamlines of this model together with isocontours of the magnitude of the velocity field. The red cross in this figure indicates the location of the so-called stagnation point where, according to \eqs{e18} and \eqref{e19}, the velocity field vanishes. From these equations we can see that it is located at $r=s$, $\theta=0$. On the other hand, from \eq{e22} we can show that the unique streamline that ends up at the stagnation point is characterized by the critical impact parameter $b_c = 2s$. Moreover, this same streamline separates the flow into two regions: all the streamlines with an impact parameter $b< b_c$ end up accreting onto the central object while those with $b> b_c$ escape to infinity. It is also interesting to note that the equation for the streamlines in \eq{e22} is independent of both the mass of the central object $M$ and the fluid pressure. 

The pressure field can now be recovered from the Bernoulli constant in \eq{e6}  as
\begin{equation}
P = P_\8 + \frac{\G M\,\rho}{r} + \rho\,v^2_\8\frac{s^2}{r^2}\left(\cos\theta-\frac{s^2}{2\,r^2} \right),  \label{e24}
\end{equation}
where we have used the velocity components  given in \eqs{e18} and \eqref{e19}.

From \eq{e24} we see that $P$ will become negative at sufficiently small radii. In order to prevent this from happening, we can take the accretor to have a finite radius $R$ and require the central object to fully enclose the region where $P\le0$. Using \eq{e24}, it is easy to see that this condition can be expressed as an upper limit on the accretion rate, namely
\begin{equation}
\dot{M} \le 4\pi\,\rho\,v_\8\,R^2\left(\sqrt{1+\frac{2\,P_\8}{\rho\,v^2_\8}+\frac{2\,\G M}{R\,v^2_\8}}-1\right).
\label{e25}
\end{equation}

Finally, we can also require that $s>R$, i.e.~that the stagnation point is located outside the central accretor. This condition, together with \eq{e25}, leads to the following upper limit for the wind speed at infinity
\begin{equation}
v_\8 < \sqrt{\frac{2}{3}\left(\frac{\G M}{R} + \frac{P_\8}{\rho}\right)}.
\label{e26}
\end{equation}

\section{Relativistic Model}
\label{Sec3}

The relativistic solution found by Petrich, Shapiro \& Teukolsky\cite{petrich88} (PST) describes a wind accreting onto a rotating black hole (Kerr spacetime) under the conditions of stationarity and irrotational flow. Similarly to the Newtonian case, an irrotational flow in general relativity can be described by a velocity potential $\Phi$ such that\cite{conventions}
\begin{equation}
 h\, u^\mu = g^{\mu\nu}\, \frac{\partial\, \Phi}{\partial x^\nu},
\label{e27}
\end{equation}
where $h= (e+P)/\rho\,\cc^2$ is the relativistic enthalpy, $e$ the relativistic internal energy density,\cite{notation} $u^\mu = \ud x^\mu/\ud \tau$ the four-velocity, and $g^{\mu\nu}$ the inverse of the spacetime metric. Substituting \eq{e27} into the continuity equation (in this case $\nabla_\mu (\rho\,u^\mu)=0$, where $\nabla_\mu$ stands for the covariant derivative) leads to
\begin{equation}
g^{\mu\nu}\nabla_\mu \left( \frac{\rho}{h}\,\frac{\partial\, \Phi}{\partial x^\nu} \right) = 0.
\label{e28}
\end{equation}

\begin{figure*}
\begin{center}
  \includegraphics[width=0.45\textwidth]{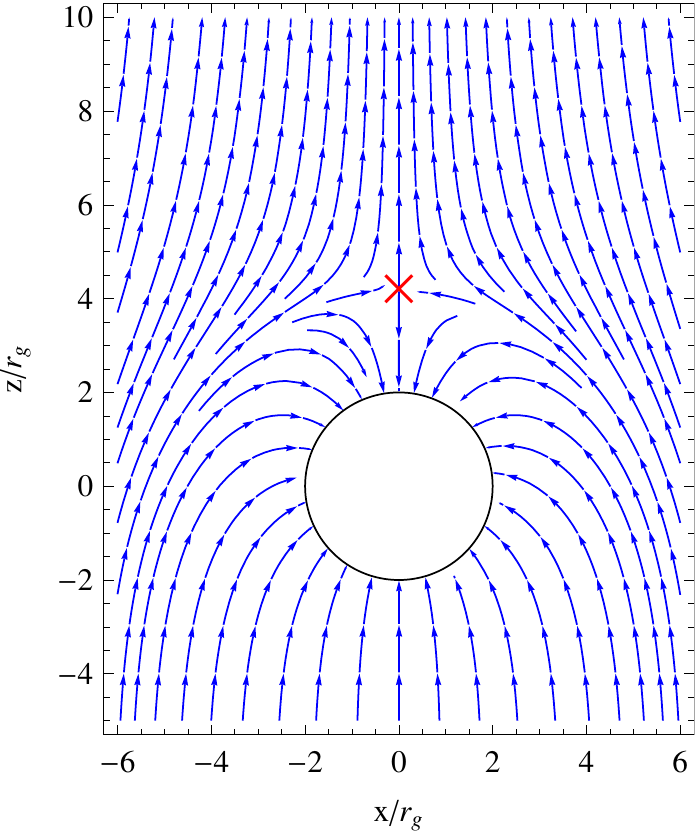}   
  \includegraphics[width=0.45\textwidth]{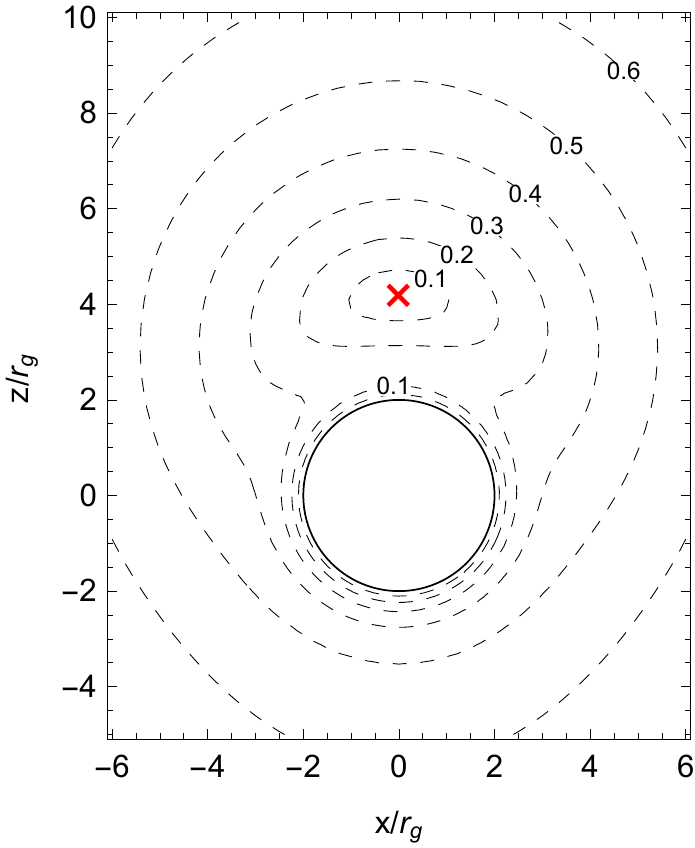}
\end{center}
 \caption{Wind accretion of the PST model for a Schwarzschild black hole. In this case we have taken $\sigma=3\,r_g$ (which corresponds to $v_\8 \simeq 0.44\,\cc$). The left-hand panel shows the streamlines as described by \eq{e40}. The right-hand panel shows the isocontour levels of the magnitude of the velocity field in \eqs{e36} and \eqref{e37}, the contour labels being expressed in units of $v_\8$. The stagnation point (located at $\theta=0$, $r=  r_g + (r_g^2+\sigma^2)^{1/2}  \simeq 4.2\,r_g$) is marked with a red cross in both panels. The black circle shows the black hole's event horizon located at the Schwarzschild radius $2\,r_g$.}
\label{f2}
\end{figure*}

In general, Eq.~\eqref{e28} is a non-linear equation in $\Phi$ except in the special case where $h\propto\rho$. Specifically, PST considered a stiff equation of state for which $P \propto \rho^2$ and $P = e$.\cite{caveat} With this choice, the fluid's sound speed is constant and equal to the speed of light $\cc$ everywhere in the fluid.\cite{soundspeed}

In the case of a non-rotating black hole (i.e. Schwarzschild spacetime), PST found the following expression for the velocity potential
\begin{equation}
\begin{split}
\Phi  = -\Gamma_\8\Big[t\,\cc^2 + 2\,\cc\,r_g\, &\ln\left(1-2\,\frac{r_g}{r}\right) \\
& - v_\8(r-r_g)\cos\theta\Big],
\end{split}
\label{e29}
\end{equation}
where $r_g = \G M /\cc^2$ is the gravitational radius and $\Gamma_\8$ is the Lorentz factor as measured at infinity 
\begin{equation}
\Gamma_\8 \equiv \left.\left(\frac{\ud t}{\ud \tau}\right)\right|_\8 = \frac{1}{\sqrt{1-v^2_\8/\cc^2}}.
\label{e30}
\end{equation}
Substituting the velocity potential $\Phi$ into \eq{e27} leads to the velocity field
\begin{align}
\rho\,\frac{\ud t}{\ud \tau} & = \Gamma_\8\,\rho_\8\left(1-2\,\frac{r_g}{r}\right)^{-1} 
\label{e31}\\
\rho\,\frac{\ud r}{\ud \tau} & = \Gamma_\8\,\rho_\8\left[v_\8\left(1-2\,\frac{r_g}{r}\right)\cos\theta - 4\,\cc\,\frac{r_g^2}{r^2}\right], 
\label{e32} \\
\rho\,\frac{\ud \theta}{\ud \tau} & = -\Gamma_\8\,\rho_\8\frac{v_\8}{r^2}(r-r_g)\sin\theta. 
\label{e33}
\end{align}

The special restrictions that arise in general relativity from demanding a regular solution across the black hole's event horizon (located at $2\,r_g$ for a Schwarzschild spacetime) imply that the PST model is characterized by the unique accretion rate
\begin{equation}
\dot{M} = 16\pi\frac{(\G M)^2}{\cc^3}\rho_\8\,\Gamma_\8.
\label{e34}
\end{equation}
This is an important difference with respect to the Newtonian solution discussed in the previous section where the accretion rate was a free parameter (only restricted by the inequality in Eq.~\ref{e23}).

Let us now define the constant
\begin{equation}
\sigma = \sqrt{\frac{\dot{M}}{4\pi\rho_\8 v_\8 \Gamma_\8}} = \frac{s}{\sqrt{\Gamma_\8}},
\label{e35}
\end{equation}
as a natural extension of the stream length scale $s$ introduced in \eq{e20}. Using this definition together with \eq{e31}, we can rewrite the velocity components in \eqs{e32} and \eqref{e33} in terms of the coordinate time $t$ as
\begin{align}
\frac{\ud r}{\ud t} & = v_\8\left(1-2\,\frac{r_g}{r}\right)\left[\left(1-2\,\frac{r_g}{r}\right)\cos\theta - \frac{\sigma^2}{r^2} \right], 
\label{e36} \\
\frac{\ud \theta}{\ud t} & = -\frac{v_\8}{r}\left(1-2\,\frac{r_g}{r}\right)\left(1-\frac{r_g}{r}\right)\sin\theta,
\label{e37}
\end{align}
and consider the non-relativistic limit, that is, the regime in which $v_\8/\cc \ll 1$ and $r_g/r\ll 1$. Within this limit, we have that $\Gamma_\8\rightarrow 1$ and $\sigma\rightarrow s$. It then follows that, within this same limit, the velocity components in \eqs{e36} and \eqref{e37} reduce to the ones corresponding to the Newtonian model in \eqs{e18} and \eqref{e19}.

\begin{figure*}
\begin{center}
  \includegraphics[width=0.4\textwidth]{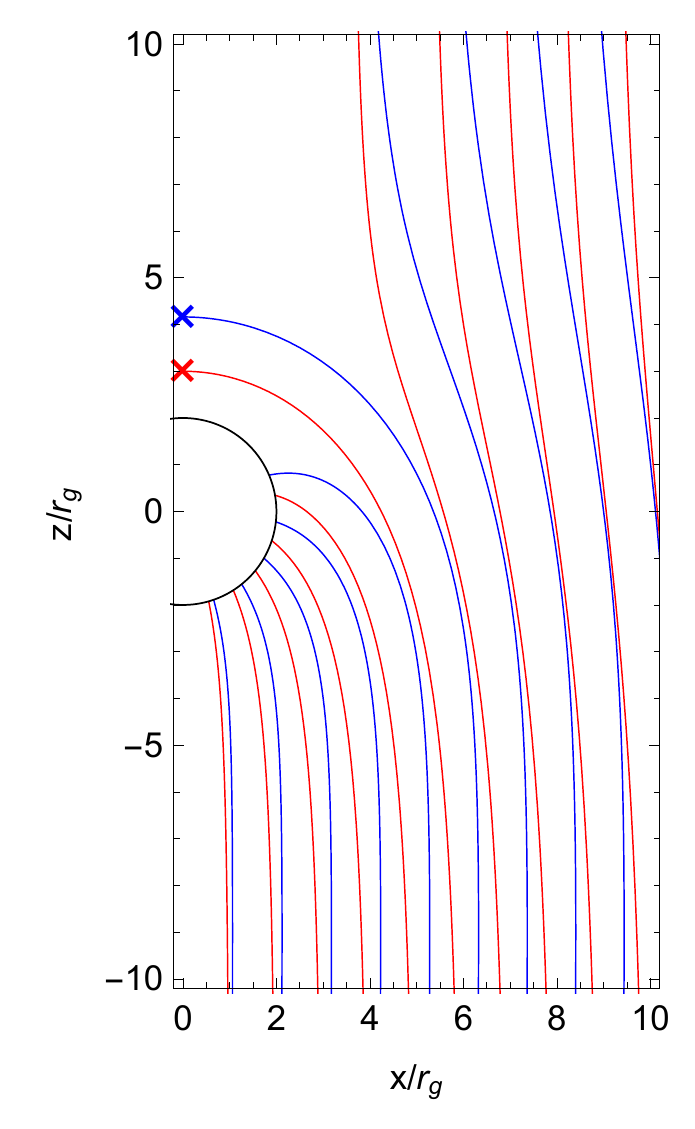}   
  \includegraphics[width=0.4\textwidth]{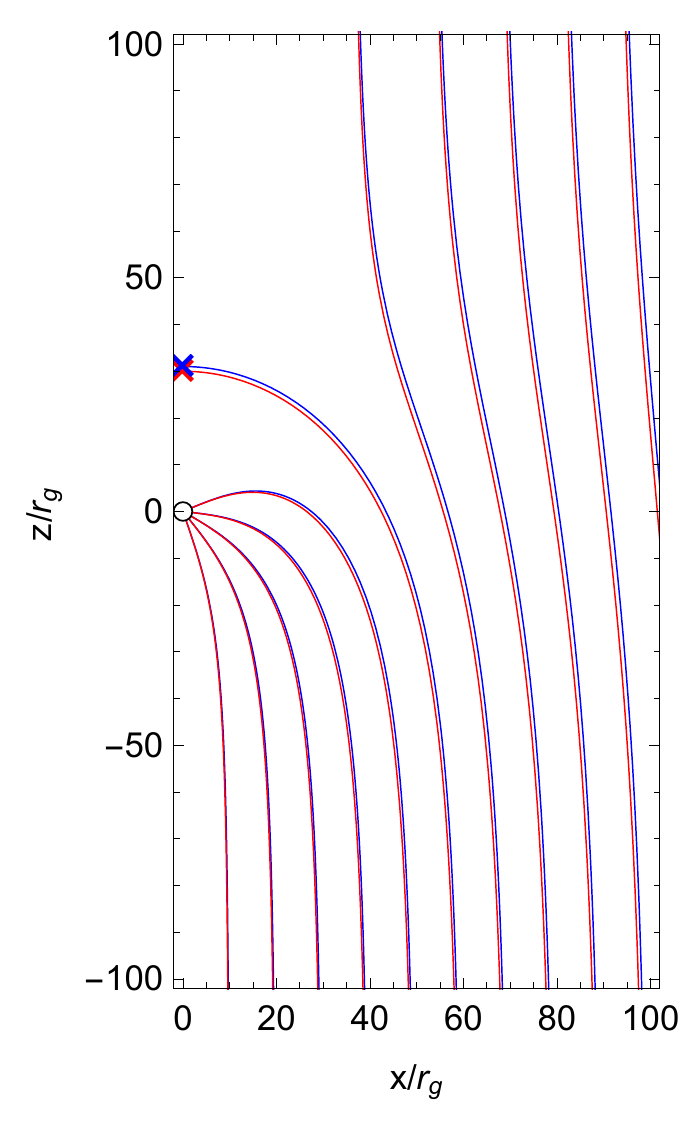}
\end{center}
 \caption{Comparison between the streamlines of the relativistic PST model (blue lines) and the Newtonian model discussed in Section~\ref{Sec2} (red lines). In the left-hand panel we have taken $s=\sigma=3\,r_g$ while, for the right-hand one we have $s=\sigma=30\,r_g$. In both panels, the black circle shows the black hole's event horizon located at the Schwarzschild radius $2\,r_g$.}
\label{f3}
\end{figure*}

In contrast to the Newtonian solution, the mass density $\rho$ in the PST model is not a constant field. An expression for $\rho$ can be found by imposing the normalization condition of the four-velocity ($g_{\mu\nu}u^\mu u^\nu=-\cc^2$) in \eqs{e31}--\eqref{e33}, which results in
\begin{equation}
\begin{split}
\rho & = \Gamma_\8\,\rho_\8\Bigg[  \left(1+2\,\frac{r_g}{r}\right) \left(1+4\,\frac{r_g^2}{r^2}\right) + 8\,\frac{v_\8}{\cc}\,\frac{r_g^2}{r^2} \cos\theta \\
& \hspace{2cm} -\frac{v^2_\8}{\cc^2}\left(1- 2\,\frac{r_g}{r}+ \frac{r_g^2}{r^2}\sin^2\theta\right) \Bigg]^{1/2}. 
\label{e38}
\end{split}
\end{equation}
From this expression it is simple to see that, in the non-relativistic limit, we have \mbox{$\rho\rightarrow \rho_\8=\co$} In other words, within this limit we recover the incompressible flow approximation on which we based our Newtonian solution.

Just as in the Newtonian case, an equation for the streamlines can be obtained by first combining \eqs{e36} and \eqref{e37} as\cite{karas93}
\begin{equation}
\frac{\ud r}{\ud \theta} = \frac{\sigma^2\csc\theta}{r-r_g} - r\left(\frac{r-2\,r_g}{r-r_g}\right)\cot\theta,
\label{e39}
\end{equation}
and then integrating this differential equation to obtain the following expression
\begin{equation}
r = r_g+\sqrt{\frac{b^2 - 2\,\sigma^2(\cos\theta+1)}{\sin^2\theta}+r_g^2}, 
\label{e40}
\end{equation}
where, as before, the integration constant $b$ corresponds to the impact parameter characterizing each individual streamline. 

From the velocity components in \eqs{e36} and \eqref{e37} we can see that the stagnation point in this case is located at $r= r_g + (r_g^2+\sigma^2)^{1/2} $, $\theta =0$. Completely analogously to the Newtonian case, from \eq{e40} it can be shown that the critical impact parameter $b_c = 2\,\sigma$ corresponds to the unique streamline ending up at the stagnation point. By combining \eqs{e34} and \eqref{e35} we obtain the following relationship between the stream length scale $\sigma$ and the wind speed at infinity
\begin{equation}
\frac{v_\8}{\cc} = 4\,\frac{r_g^2}{\sigma^2},
\label{e41}
\end{equation}
from which it can be shown that even in the limit $v_\8\rightarrow\cc$, the stagnation point reaches a minimum radius of $r\simeq 3.2\,r_g$, i.e. it is always located outside the event horizon.

Note that the velocity field in \eqs{e36} and \eqref{e37} also vanishes at the  event horizon $r=2\,r_g$. This is only a coordinate effect related to the fact that the light cones close onto themselves when described in terms of the coordinate time $t$, which implies that $t$ is not well suited for describing physical processes close to the event horizon. Indeed, from \eqs{e31}--\eqref{e33}, it is simple to see that the velocity components $\ud r/\ud \tau$ and $\ud \theta/\ud \tau$ do not show this behavior.

Contrary to the Newtonian model where the resulting flow was described by only one characteristic length scale, namely the stream length scale $s$, the relativistic model is characterized by two length scales: the stream length scale $\sigma$ and the gravitational radius $r_g$. In Figure~\ref{f2} we show the streamlines and isocontour levels of the velocity field of the PST model for the particular case $\sigma = 3\,r_g$ which, from \eq{e41}, corresponds to a wind speed at infinity of $v_\8 \simeq 0.44\,\cc$.

In order to facilitate the comparison between the Newtonian and the PST models, in Figure~\ref{f3} we show the corresponding streamlines side by side for two cases: $s=\sigma=3\,r_g$ and $s=\sigma=30\,r_g$. 
From \eq{e41} it is simple to see that $\sigma \gg r_g$ implies $v_\8 \ll \cc$. Therefore, as the ratio $\sigma/r_g$ grows, the non-relativistic limit should be recovered, just as we can confirm by comparing the left and right panels of this figure.

\section{Summary}

In this paper, we have presented a simple analytic model of wind accretion for an incompressible fluid falling onto a massive gravitating object. We have shown that this model corresponds to the Newtonian limit of the relativistic solution of wind accretion onto a black hole found by Petrich, Shapiro \& Teukolsky\cite{petrich88} (PST). The fluid in the PST model obeys a stiff equation of state $P=e$ for which the sound speed is constant everywhere and equal to the speed of light. In hindsight, it is not surprising that the Newtonian limit of such a fluid corresponds precisely to an incompressible fluid for which the sound speed is, formally, equal to infinity.

The Newtonian model features only one characteristic length scale: the stream length scale $s=(\dot{M}/4\pi\rho\,v_\8)^{1/2}$. In addition to this the PST model is also characterized by the gravitational radius $r_g = \G M/\cc^2$. The existence of these two characteristic length scales leads to a richer variety of flow morphologies than for the Newtonian model (see Figure~\ref{f3}). The Newtonian model is recovered in the limit $s\gg r_g$ which naturally coincides with the non-relativistic limit $v_\8 \ll \cc$.

Another difference between the Newtonian and relativistic models is that, in the former, the accretion rate $\dot{M}$ is a free parameter while, in the latter, it has to have the fixed value $\dot{M}=16\pi(\G M)^2\rho_\8\,\Gamma_\8/\cc^3$ in order to guarantee a regular solution across the black hole's event horizon.

The Newtonian model presented in this article can be used as an illustrative example during a gas dynamics course. Moreover, it should be useful as a benchmark for testing Newtonian hydrodynamics codes.
\vspace{12pt}

\begin{acknowledgments}

I gratefully acknowledge John Miller, Sergio Mendoza and Olivier Sarbach for helpful comments and suggestions.  This work was supported by the CONACyT grants 290941 and 291113.

\end{acknowledgments}


\begin{thebibliography}{10}
\newcommand{\enquote}[1]{``#1''}
\providecommand{\url}[1]{\texttt{#1}}
\providecommand{\urlprefix}{URL }
\expandafter\ifx\csname urlstyle\endcsname\relax
  \providecommand{\doi}[1]{doi:\discretionary{}{}{}#1}\else
  \providecommand{\doi}{doi:\discretionary{}{}{}\begingroup
  \urlstyle{rm}\Url}\fi
\providecommand{\eprint}[2][]{\url{#2}}

\bibitem{hoylely}
F.~{Hoyle} and A.~{Lyttleton}, \enquote{The effect of interstellar matter on
  climatic variation,} Mathematical Proceedings of the Cambridge Philosophical
  Society, \textbf{35}, 405--415 (1939).

\bibitem{hoyle}
H.~{Bondi} and F.~{Hoyle}, \enquote{{On the mechanism of accretion by stars},}
  Monthly Notices of the Royal Astronomical Society, \textbf{104}, 273 (1944).
  
\bibitem{bondi52}
H.~{Bondi}, \enquote{On spherically symmetrical accretion,}
  Monthly Notices of the Royal Astronomical Society, \textbf{112}, 195 (1952).
  
\bibitem{michel72}
F.~C.~{Michel}, \enquote{Accretion of Matter by Condensed Objects,}
Astrophysics and Space Science, \textbf{15}, 153--160 (1972).

\bibitem{hunt71}
R.~Hunt, \enquote{A fluid dynamical study of the accretion process,}
  Monthly Notices of the Royal Astronomical Society, \textbf{154}, 141 (1971).

\bibitem{bisnovatyi}
G.~S. {Bisnovatyi-Kogan}, I.~M. {Kazhdan}, A.~A. {Klypin}, A.~E. {Lutskii}, and
  N.~I. {Shakura}, \enquote{{Accretion onto a rapidly moving gravitating
  center},} Astronomicheskii Zhurnal, \textbf{56}, 359--367 (1979).

\bibitem{petrich89}
L.~I. {Petrich}, S.~L. {Shapiro}, R.~F. {Stark}, and S.~A. {Teukolsky},
  \enquote{{Accretion onto a moving black hole - A fully relativistic
  treatment},} Astrophysical Journal, \textbf{336}, 313--349 (1989).

\bibitem{matsuda91}
T.~{Matsuda}, N.~{Sekino}, K.~{Sawada}, E.~{Shima}, M.~{Livio}, U.~{Anzer}, and
  G.~{Boerner}, \enquote{{On the stability of wind accretion},} Astronomy and
  Astrophysics, \textbf{248}, 301--314 (1991).

\bibitem{font98a}
J.~A. {Font} and J.~M. {Ib{\'a}{\~n}ez}, \enquote{{A Numerical Study of
  Relativistic Bondi-Hoyle Accretion onto a Moving Black Hole: Axisymmetric
  Computations in a Schwarzschild Background},} Astrophysical Journal,
  \textbf{494}, 297--316 (1998).

\bibitem{foglizzo05}
T.~{Foglizzo}, P.~{Galletti}, and M.~{Ruffert}, \enquote{{A fresh look at the
  unstable simulations of Bondi-Hoyle-Lyttleton accretion},} Astronomy and
  Astrophysics, \textbf{435}, 397--411 (2005).

\bibitem{penner11}
A.~J. {Penner}, \enquote{{General relativistic magnetohydrodynamic Bondi-Hoyle
  accretion},} Monthly Notices of the Royal Astronomical Society, \textbf{414},
  1467--1482 (2011).

\bibitem{zanotti11}
O.~{Zanotti}, C.~{Roedig}, L.~{Rezzolla}, and L.~{Del Zanna}, \enquote{{General
  relativistic radiation hydrodynamics of accretion flows - I. Bondi-Hoyle
  accretion},} Monthly Notices of the Royal Astronomical Society, \textbf{417},
  2899--2915 (2011).

\bibitem{cruz12}
A.~{Cruz-Osorio}, F.~D. {Lora-Clavijo}, and F.~S. {Guzm{\'a}n}, \enquote{{Is
  the flip-flop behaviour of accretion shock cones on to black holes an effect
  of coordinates?}} Monthly Notices of the Royal Astronomical Society,
  \textbf{426}, 732--738 (2012).

\bibitem{blondin12}
J.~M. {Blondin} and E.~{Raymer}, \enquote{{Hoyle-Lyttleton Accretion in Three
  Dimensions},} Astrophysical Journal, \textbf{752}, 30 (2012).

\bibitem{LG13}
F.~D. {Lora-Clavijo} and F.~S. {Guzm{\'a}n}, \enquote{{Axisymmetric Bondi-Hoyle
  accretion on to a Schwarzschild black hole: shock cone vibrations},} Monthly
  Notices of the Royal Astronomical Society, \textbf{429}, 3144--3154 (2013).

\bibitem{livio91}
M.~{Livio}, \enquote{{On accretion by compact objects from a stellar wind},} in
  \emph{Frontier Objects in Astrophysics and Particle Physics.}, edited by
  F.~{Giovannelli} and G.~{Mannocchi} (1991), pp. 67--80.

\bibitem{edgar04}
R.~{Edgar}, \enquote{{A review of Bondi-Hoyle-Lyttleton accretion},} New
  Astronomy Review, \textbf{48}, 843--859 (2004).

\bibitem{petrich88}
L.~I. {Petrich}, S.~L. {Shapiro}, and S.~A. {Teukolsky}, \enquote{{Accretion
  onto a moving black hole - an exact solution},} Physical Review Letters,
  \textbf{60}, 1781--1784 (1988).
  
\bibitem{conventions}
Greek indices run over spacetime components, e.g.~$x^\mu=(\cc\,t,\,r,\,\theta,\,\phi)$. We adopt Einstein's summation convention over repeated indices.

\bibitem{notation}
Note that in the PST paper\cite{petrich88} the authors use the symbol $\rho$ to denote the relativistic internal energy density and work with the baryon number density $n$. For an average baryonic rest
mass $m$, $n$ is related to the mass density by $\rho = m\,n$.

\bibitem{caveat}
The relativistic energy density is defined as $e = \rho(u +\cc^2)$, where $u$ is the non-relativistic internal energy (per unit mass). For a polytropic equation of state $P=K\,\rho^2$ one has $u=K\,\rho$. Thus, the equations $P \propto \rho^2$ and $P = e$ are incompatible unless the rest-mass energy is negligible compared to the internal energy, i.e.~$u\gg\cc^2$. These two equations will also be compatible for the case of a massless scalar field, in which case the number density $n$ plays the role of a comoving marker.

\bibitem{soundspeed}
The relativistic expression for the sound speed is $a=\cc \sqrt{\partial P/\partial e|_s}$ as opposed to $a= \sqrt{\partial P/\partial \rho|_s}$ in non-relativistic physics.

\bibitem{karas93}
V.~{Karas} and R.~{Mucha}, \enquote{{Accretion onto a rotating compact object
  in general relativity},} American Journal of Physics, \textbf{61}, 825--828
  (1993).

\end{thebibliography}

\end{document}